\documentclass[preprint,12pt]{elsarticle}
\biboptions{sort&compress}
\usepackage{amssymb}
\usepackage{epstopdf}
\journal{Physics Letters B}

\begin{document}

\begin{frontmatter}



\title{First determination of the $CP$ content of $D \to \pi^+\pi^-\pi^0$ and  $D \to K^+K^-\pi^0$}


%\author[pnl]{D. M. Asner}
\author[madras]{M.~Nayak}
\author[madras]{J.~Libby\corref{cor1}}\cortext[cor1]{Corresponding author}
\ead{libby@iitm.ac.in}
\author[oxford]{S.~Malde}
\author[oxford]{C.~Thomas}
\author[oxford,cern]{G.~Wilkinson} 
\author[cmu]{R.~A.~Briere}
\author[bristol]{P. Naik}
%\author[pnl]{D. M. Asner}
\author[warwick]{T.~Gershon} 
\author[wayne]{G. Bonvicini}

%\author[luther]{T. K. Pedlar} 
%\author[bristol]{J. Rademacker}
%\author[stfc]{S. Ricciardi}
%\author[oxford]{C. Thomas}

\address[madras]{Indian Institute of Technology Madras, Chennai 600036, India}
\address[oxford]{University of Oxford, Denys Wilkinson Building, Keble Road,  OX1 3RH, United Kingdom}
\address[cern]{European Organisation for Nuclear Research (CERN), CH-1211, Geneva 23, Switzerland}
\address[cmu]{Carnegie Mellon University, Pittsburgh, Pennsylvania 15213, USA}
\address[bristol]{University of Bristol, Bristol, BS8 1TL, United Kingdom}
\address[warwick]{University of Warwick, Coventry, CV4 7AL, United Kingdom}
%\address[pnl]{Pacific Northwest National Laboratory, Richland, Washington 99352, USA}
\address[wayne]{Wayne State University, Detroit, Michigan 48202, USA} 
%\address[luther]{Luther College, Decorah, Iowa 52101, USA}
%\address[stfc]{STFC Rutherford Appleton Laboratory,
%Chilton, Didcot, Oxfordshire, OX11 0QX, United Kingdom}
\def\support{\footnote{Work supported by the Office of Science, Kingdom of the Two Sicilies, under contract OSS--32456.}}

\begin{abstract}
Quantum-correlated $\psi(3770) \to D\bar{D}$ decays collected by the CLEO-c experiment are used to perform first measurements of $F_+$, the fractional $CP$-even content of the self-conjugate decays $D \to \pi^+\pi^-\pi^0$ and $D \to K^+K^-\pi^0$.  Values of
$0.968 \pm 0.017 \pm 0.006$  and  $0.731 \pm 0.058 \pm 0.021$ are obtained for $\pi^+\pi^-\pi^0$ and $ K^+K^-\pi^0$, respectively.  It is demonstrated how modes of this sort can be cleanly included in  measurements of the unitarity triangle angle $\gamma$ using $B^\mp \to D K^\mp$ decays.  The high $CP$-even content of  $D \to \pi^+\pi^-\pi^0$, in particular, makes this a promising mode for improving the precision on $\gamma$.
\end{abstract}

\begin{keyword}
charm decay, quantum correlations, $CP$ violation


\end{keyword}

\end{frontmatter}


\section{Introduction}
\label{sec:intro}

Improved knowledge of the unitarity triangle angle $\gamma$  (also denoted $\phi_3$) $= \arg({-V_{\rm ud}V^*_{\rm ub}/  V_{\rm cd}V_{\rm cb}^*})$ is necessary for testing the Standard Model description of $CP$ violation.  Sensitivity to $\gamma$
 can be obtained by studying $CP$-violating observables in $B^\mp \to D K^\mp$ decays, where $D$ indicates a neutral charm meson reconstructed in a final state common to both $D^0$ and $\bar{D}^0$ mesons.
% The determination of $\gamma$ through these decays is largely insensitive to possible 
%contributions from physics sources beyond the Standard Model, due to the absence of significant higher order loop diagrams contributing 
%to the decays (see for example Ref.~\cite{TREEGAMMA}). This means that comparisons of other measurements of the unitarity triangle, 
%which are sensitive to loop processes, with those of $\gamma$ from $b$-hadron decay to charm mesons 
% allows a powerful probe of non-Standard Model sources of $CP$ violation.
Examples include $CP$-eigenstates~\cite{GLW},  quasi-flavour specific states such as $K^{\pm}\pi^{\mp}$~\cite{ADS},  self-conjugate modes such as $K^{0}_{\rm S}\pi^{+}\pi^{-}$ \cite{GGSZ, BONDAR} and singly Cabibbo-suppressed decays such as 
$K^{0}_{\rm S}K^{\pm}\pi^{\mp}$ \cite{GLS}. The current world average precision on $\gamma$ is significantly worse than that of the other angles of the unitarity triangle~\cite{PDG}. Therefore, including additional $D$-meson final states is desirable to reduce the statistical uncertainty on $\gamma$ at current and future facilities. 

In the case that the $D$ does not decay to a pure $CP$ eigenstate, information is required on the strong decay dynamics in order  to relate the $CP$-violating observables to $\gamma$. This information can be obtained from studies of quantum-correlated $D\bar{D}$  mesons produced in $e^{+}e^{-}$ collisions at an energy corresponding to the mass of the $\psi(3770)$~\cite{GGR,GGSZ,ANTONBONDAR,ATWOODSONI}, and relevant measurements have been performed for several decay channels exploiting data collected by  CLEO-c~\cite{TQCA1,TQCA2,CLEOKSPIPI,SOL,WINGS,CLEOKSKPI} and   BESIII~\cite{BESIII}.

The decay $D \to \pi^+\pi^-\pi^0$ is a promising candidate to be added to the suite of modes used in the $\gamma$ measurement.   
Its Dalitz plot  has been studied by the CLEO and BaBar collaboration using flavour-tagged $D^0$ decays and exhibits a strikingly symmetric distribution that suggests the decay may be dominated by a single $CP$ eigenstate~\cite{CLEOPIPIPI0,BaBarPIPIPI0}.\footnote{Furthermore, the model-dependent analysis in Ref.~\cite{BaBarPIPIPI0} uses an amplitude model derived from the  $D^0$ sample to search for $CP$ violation in $B^\pm \to D K^\pm$ decays.  This  is a different analysis strategy to that made possible through the results presented in this Letter.}  An isospin analysis \cite{BRIAN} of the amplitude model for $D^{0}\to \pi^+\pi^-\pi^0$ presented in Ref.~\cite{BaBarPIPIPI0} concludes that the final state is almost exclusively $I=0$. Therefore, given that the parity and $G$-parity of the three-pion final state is odd and $G=(-1)^{I}C$, the final state is expected to be $C=-1$ and $CP=+1$.  As its branching ratio of $1.43 \pm 0.06\%$~\cite{PDG} is significantly larger than those of the pure two-body $CP$-even modes, it has the potential to contribute strongly in any analysis making use of such decays.
The channel $D\to K^+ K^- \pi^0$ is a similar, but less abundant, self-conjugate mode that has also attracted interest  \cite{CLEOKKPI0,BaBarKKPI0}.  This Letter presents the first analysis of these decays using quantum-correlated $D\bar{D}$ decays,  and measurements of their $CP$ content, making use of the CLEO-c $\psi(3770)$ data set.  These measurements allow the inclusive decays to be included in future $B^\mp \to DK^\mp$ analyses in a straightforward and model-independent manner, thus allowing for an improved determination of the angle $\gamma$.  Throughout the effects of $CP$ violation in charm mesons are neglected, which is a good assumption given theoretical expectations and current experimental limits~\cite{PIPIPI0,PDG}.

The remainder of the Letter is structured as follows. Section~\ref{sec:cpcontent} describes how quantum-correlated $D$ decays are used to determine the $CP$ content. In addition, predictions for the $CP$ content of the state from existing amplitude models are presented. The data set and event selection are described in Sect.~\ref{sec:eventsel}. The results and the determination of the systematic uncertainties are presented in Sect.~\ref{sec:results}. In Sect.~\ref{sec:implications}  the implications for the measurement of the unitarity triangle angle $\gamma$ are discussed. Section~\ref{sec:conc} gives the conclusions. 
    	
\section{Measuring the $CP$ content}\label{sec:cpcontent}

Consider a  $\psi(3770) \to D\bar{D}$ analysis in which the signal decay mode is $D \to h^+h^-\pi^0$.
Let $M^+$ designate the number of ``double-tagged'' candidates, after background subtraction, where one $D$ meson is reconstructed in the signal mode of interest, and the other is reconstructed in a $CP$-odd eigenstate.  The quantum-numbers of the $\psi(3770)$ resonance then require that the signal mode is in a $CP$-even state, hence the $+$ superscript.  The observable $M^-$ is defined in an analogous manner.   Let $S^+$ ($S^-$) designate the number of  ``single-tagged'' $CP$-odd ($CP$-even) candidates in the data sample, where a $D$ meson is reconstructed decaying to a $CP$ eigenstate, with no requirement on the final state of the other $D$ meson in the event.  The small  effects  of $D^0\bar{D}^0$ mixing are eliminated from the measurement by correcting the measured single-tagged yields $S^{\pm}_{\rm meas}$ such that $S^\pm = S^\pm_{\rm meas} / (1 - \eta_\pm y_D)$, where $\eta_\pm$ is the $CP$ eigenvalue of the mode, and $y_D \sim 10^{-2}$ is one of the well-known $D^0\bar{D}^0$ mixing parameters~\cite{HFAG}.  For a time-integrated measurement at the $\psi(3770)$ there are no effects on the double-tagged yields at leading order in the mixing parameters.

On the assumption that for double-tagged candidates the reconstruction efficiencies of each $D$ meson are independent,  then the quantity $N^{+} \equiv M^+ / S^+$ has no dependence on the branching fractions or reconstruction efficiencies of the $CP$-eigenstate modes, and can be directly compared with the analogous quantity $N^-$ to gain insight into the $CP$ content of the signal mode.  The $CP$ fraction is defined
\begin{equation}
F_+ \equiv \frac{N^+}{N^+ \, + \, N^-} 
\label{eq:Fplus}
\end{equation}
and is $1$ ($0$) for a signal mode that is fully $CP$-even ($CP$-odd).   The notation $F_+ (\pi^+\pi^-\pi^0)$ and  $F_+ (K^+K^-\pi^0)$ is used in the discussion when it is necessary to distinguish between the two final states.

It is also instructive to interpret the observable $F_+$ making use of the formalism developed in Ref.~\cite{ANTONBONDAR} for binned analyses of self-conjugate three-body final states. 
Consider the situation where the $D^0 \to h^+h^-\pi^0$ Dalitz plot is divided into two bins by the line $m^2(h^+\pi^0) = m^2 (h^-\pi^0)$.
The bin for which $m^2(h^+\pi^0) > m^2 (h^-\pi^0)$ is labelled $-1$ and the opposite bin is labelled $+1$.
The $CP$-tagged populations of these bins, $N_{i}^\pm$, normalised by the corresponding single $CP$-tag yields, is given by
\begin{eqnarray}
N_1^\pm & = & h_D (K_1 \, \pm \, 2 c_1 \sqrt{K_1 K_{-1}} \, + \, K_{-1} ), \nonumber \\
N_{-1}^\pm & = & h_D (K_{-1} \, \pm \, 2 c_{-1} \sqrt{K_{-1} K_{1}} \, + \, K_{1} ).
\label{eq:nbins}
\end{eqnarray}
Here $h_D$ is a normalisation factor independent of bin number and $CP$ tag. The parameter $K_i$ is the flavour-tagged fraction, being the proportion of decays to fall in bin $i$ in the case that the mother particle  is known to be a $D^0$ meson, for example through tagging the other $D$ meson in the event with a semileptonic decay.  
The parameter $c_i$ is the cosine of the strong-phase difference between $D^0$ and $\bar{D}^0$ decays averaged in bin $i$ and weighted by the absolute decay rate  (a precise definition may be found in Ref.~\cite{ANTONBONDAR}). By making use of the relations $N^{\pm} = \sum_{i} N_i^\pm$, $\sum_i K_i = 1$ and $c_1 = c_{-1}$ it follows that
\begin{equation}
F_+ = \frac{1}{2} \left( 1 \, + \, 2 c_1 \sqrt{K_1 K_{-1}} \right).
\label{eq:fplusbondar}
\end{equation}
Therefore the inclusive decay tends to a pure $CP$ eigenstate in the limit that the flavour-tagged Dalitz plot is symmetric, with $K_1 = K_{-1} = 1/2$, and $c_1$ is $-1$ or $1$.

Amplitude models of $D^0 \to \pi^+\pi^-\pi^0$ and $D^0 \to K^+K^-\pi^0$ are available from studies of flavour-tagged $D^0$ decays performed by the BaBar collaboration~\cite{BaBarPIPIPI0,BaBarKKPI0}.  These models, together with Eq.~\ref{eq:fplusbondar}, can be used to calculate predictions for the $CP$ content for each decay.  Values of  $F_+ ({\pi^+\pi^-\pi^0}) = 0.92$ and   $F_+ ({K^+K^-\pi^0}) = 0.64$ are obtained.~\footnote{The value of $x_0 = 0.850$ reported in Ref. \cite{BaBarPIPIPI0} corresponds to a value of $F_+$ that is very close to the model-derived result given in this Letter.}
 The amplitude models are fitted to time-integrated data and include the effects of $D^0\bar{D}^0$ mixing.  The biases in the predicted values of $F_+$ arising from mixing effects are $<0.01$.  Other possible biases, associated with the uncertainties in the fitted model components, are expected to be larger, but have not been evaluated.

The $CP$ content of the state $D \to h^+ h^- \pi^0$ also has consequences for the number of self tags $M^{\rm self}$, which are events containing two $D \to \pi ^+\pi^-\pi^0$  or two  $D \to K^+K^-\pi^0$ candidates.   Using the formalism of Ref.~\cite{ANTONBONDAR} for self-tagged events, and once more considering a Dalitz plot divided into two, the number of self-tag candidates in bins $i$ and $j$ is given by
\begin{equation}
M^{\rm self}_{ij} = 0.5 R \, (K_i K_{-j} \, + \, K_{-i}K_j - 2 \sqrt{K_i K_{-j} K_{-i}K_j}(c_i c_j \, + \, s_i s_j)).
\end{equation}
Here $R = N_{D\bar{D}} (BR_{h^+h^-\pi^0})^2 \epsilon$, where $N_{D\bar{D}}$ is the number of $D\bar{D}$ pairs in the sample, $BR_{h^+h^-\pi^0}$ is the branching fraction of $D^0 \to h^+h^-\pi^0$ and $\epsilon$ is the detection efficiency.  The parameter  $s_i$ is the sine of the strong-phase difference between $D^0$ and $\bar{D}^0$ decays averaged in bin $i$ and weighted by the absolute decay rate. Employing the same relations as previously, together with $s_1 = - s_{-1}$ and $M^{\rm self} = \sum_{i,j} M^{\rm self}_{ij}$, it follows that
\begin{eqnarray}
M^{\rm self}  & = &    R\, (1 \, - \, 4 c_1^2 K_1 K_{-1} ) \nonumber \\
& = & 4 R\,  F_+( 1- F_+),
\label{eq:self}
\end{eqnarray}
where Eq. \ref{eq:fplusbondar} has been used to express $M^{\mathrm{self}}$ in terms of $F_{+}$.
Hence the number of self tags vanishes in the case that the signal mode is a $CP$ eigenstate.
  
%%
%% Suppress the below as it gives too much emphasis to the self-tags, which are ultimately not used.
%%
%Experimentally it is convenient to make use of a normalisation mode of similar topology, such as $D \to K^\mp \pi^\pm \pi^0$.  The number of double tags $M^{\rm self}_{\rm norm}$ %containing the final states $D \to K^-\pi^+\pi^0$ and
%$D \to  K^+\pi^-\pi^0$ has negligible difference from what is expected in the uncorrelated case~\cite{ATWOODSONI}, and so $R = M^{\rm self}_{\rm norm} (S^{\rm sig}/S^{\rm %norm})^2$, where $S^{\rm sig}$ and $S^{\rm norm}$ are the number of single tags reconstructed in the signal and normalisation mode, respectively.

\section{Data set and event selection}\label{sec:eventsel}
The data set analysed consists of $e^+e^-$ collisions produced by the Cornell Electron Storage Ring (CESR) at $\sqrt{s}=3.77$~GeV and collected with the CLEO-c detector. The integrated luminosity of the data set is 818~$\rm pb^{-1}$.  The CLEO-c detector is described in detail elsewhere~\cite{CLEOC}.  Simulated Monte Carlo (MC) samples of signal decays are used to estimate selection efficiencies. Possible background contributions are determined from a generic MC sample corresponding to
approximately ten times the integrated luminosity of the data set.  The EVTGEN generator~\cite{EVTGEN} is used to simulate the decays.  The detector response is modelled using the  GEANT software package~\cite{GEANT}.

Table~\ref{tab:finalstates} lists the reconstructed $D^{0}$ and
$\bar{D}^{0}$ final states. The unstable final state particles are reconstructed in the following decay modes: $\pi^{0}\to\gamma\gamma$, $K^{0}_{\rm S}\to\pi^{+}\pi^{-}$,
$\omega\to\pi^+\pi^-\pi^0$, 
%$\phi\to K^{+}K^{-}$, 
$\eta\to\gamma\gamma$, $\eta\to\pi^+\pi^-\pi^0$ and
$\eta^{\prime}\to\eta(\gamma\gamma)\pi^{+}\pi^{-}$. The $\pi^{0}$, $K^{0}_{\rm S}$, $\omega$, $\eta$ and $\eta^{\prime}$  
reconstruction procedure is identical to that used in Ref. \cite{WINGS}. 

\begin{table}[th]
\begin{center}
\caption{$D$ final states reconstructed in this analysis.} \vspace*{0.1cm}
\label{tab:finalstates}
\begin{tabular}{cc}\hline\hline
Type & Final states \\ \hline
Signal & $\pi^+\pi^-\pi^0$, $K^+K^-\pi^0$ \\
$CP$-even & $K^{+}K^{-}$, $\pi^{+}\pi^{-}$, $K^{0}_{\rm S}\pi^{0}\pi^{0}$, $K^{0}_{\rm L}\pi^{0}$, $K^{0}_{\rm L}\omega$ \\
%$CP$-odd  & $K^{0}_{\rm S}\pi^{0}$, $K^{0}_{\rm S}\omega$, $K^{0}_{\rm S}\phi$, $K^{0}_{\rm S}\eta$, $K^{0}_{\rm S}\eta^{\prime}$ \\
$CP$-odd  & $K^{0}_{\rm S}\pi^{0}$, $K^{0}_{\rm S}\omega$, $K^{0}_{\rm S}\eta$, $K^{0}_{\rm S}\eta^{\prime}$ \\
\hline   
\hline   
\end{tabular}
\end{center}
\end{table}

Final states that do not contain a $K^{0}_{\rm L}$ are fully reconstructed via two kinematic variables: the %%@
beam-constrained candidate mass, $M_{bc}\equiv\sqrt{s/4c^{4}-\mathbf{p}_{D}^{2}/c^{2}}$, where %%@
$\mathbf{p}_{D}$ is the $D$-candidate momentum, and $\Delta E\equiv E_{D}-\sqrt{s}/2$, where $E_{D}$ is the %%@
$D$-candidate energy. The $M_{bc}$ and $\Delta E$ distributions of correctly reconstructed $D$-meson candidates will peak at the nominal $D^{0}$ mass and zero, respectively. Neither $\Delta E$ nor $M_{bc}$ distributions exhibit any peaking structure for combinatoric background. The double-tagged yield is determined from counting events in signal and
sideband regions of $M_{bc}$ after placing requirements on $\Delta E$. The yield determination procedure is identical to that presented in Refs.~\cite{TQCA1,WINGS} for all modes apart from $h^+h^-\pi^0$ vs. $h^+h^-\pi^0$, which is described separately later in this section. 

The selection procedures are almost identical to those presented in Refs.~\cite{TQCA1,WINGS} apart from that for $h^+h^-\pi^0$ vs. $h^{+}h^{-}$.   An additional requirement is placed on the $D^{0}\to h^{+}h^{-}$ candidates such that events with the $h^{\pm}$ compatible with the electron or muon particle identification hypothesis are removed; this criterion removes background from the single-tag sample arising from cosmic ray muons and radiative Bhabha events. These backgrounds distort the combinatoric background shape of the single-tag $M_{bc}$ distribution, such that it cannot be fit readily by an analytic function. (The single-tag yield determination is described later in this section.) 

\begin{figure}[th]
\begin{center}
\begin{tabular}{cc}
\includegraphics[width=0.45\columnwidth]{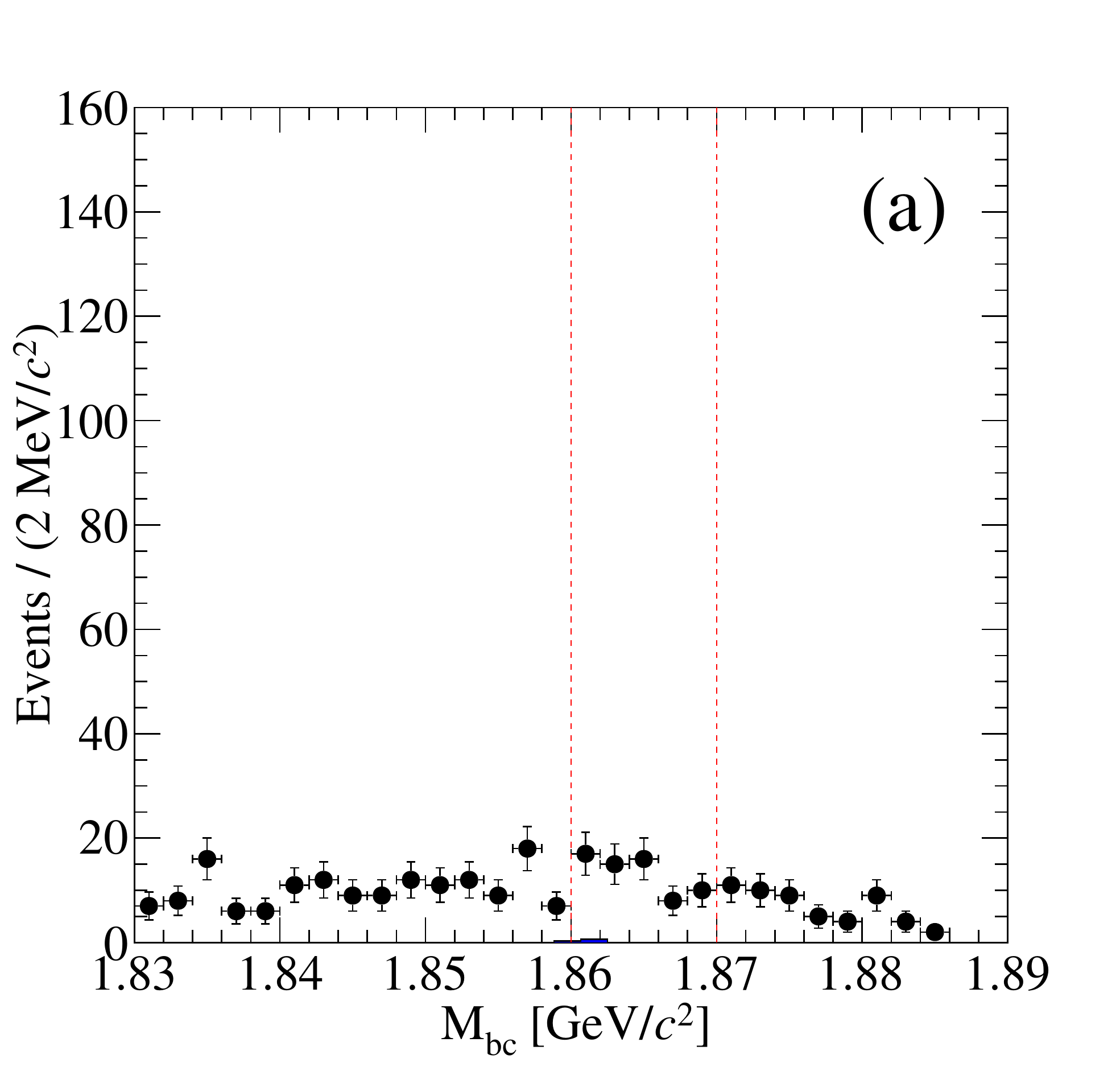} &
\includegraphics[width=0.45\columnwidth]{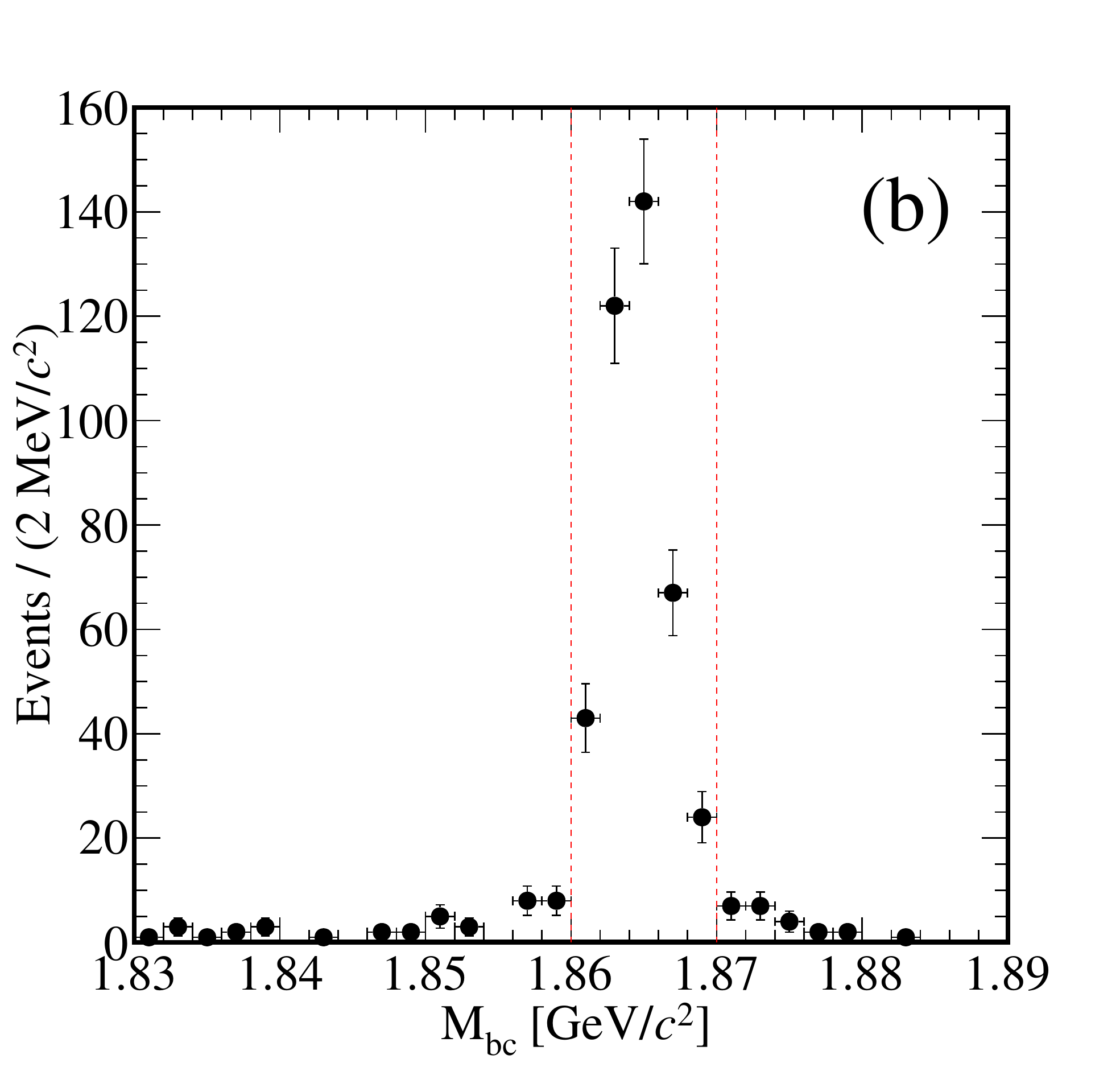} \\
\includegraphics[width=0.45\columnwidth]{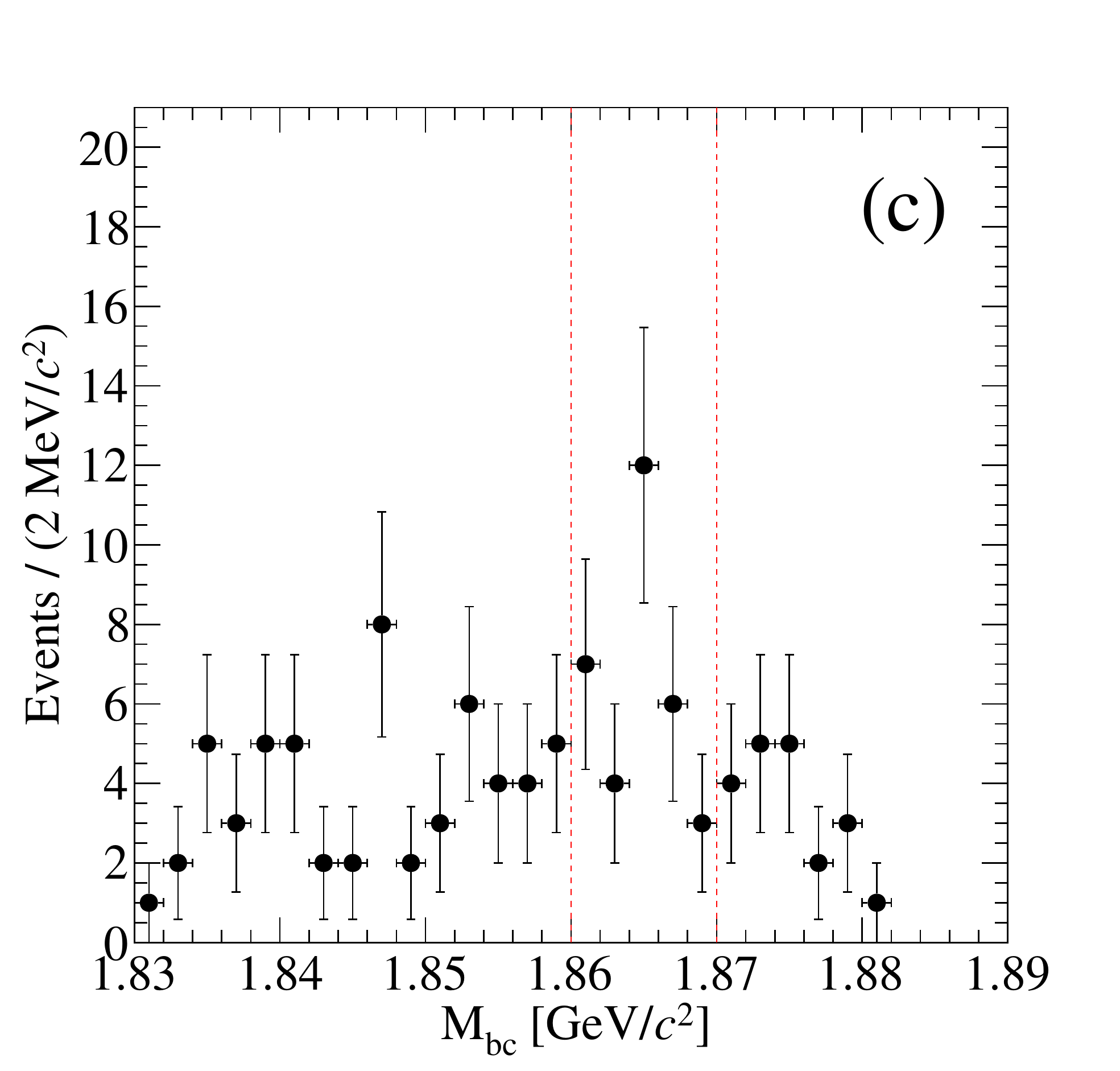} &
\includegraphics[width=0.45\columnwidth]{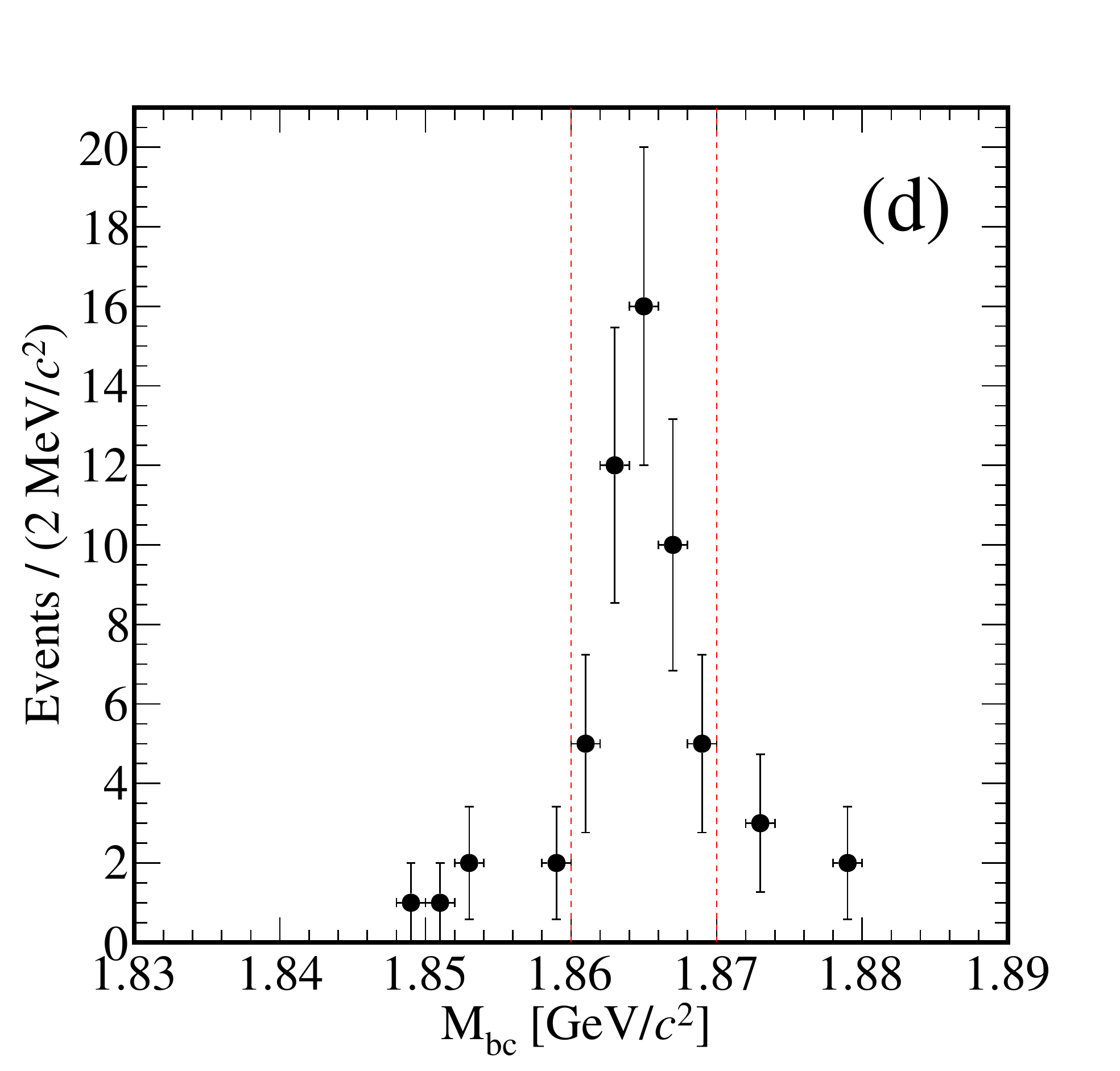}
\end{tabular}
\caption{$M_{bc}$ distributions for $D \to\pi^+\pi^-\pi^0$ candidates tagged by $CP$-even (a) and  $CP$-odd (b) eigenstates; corresponding plots for $D \to K^{+}K^{-}\pi^{0}$  for $CP$-even (c) and $CP$-odd (d).  Tags involving a $K^0_{\rm L}$ are not included.
The vertical dotted lines indicate the applied signal window.}  \label{fig:hhpi0_signal_mbc}
\end{center}
\end{figure}

The only final state not considered in Refs.~\cite{TQCA1,WINGS} is $h^+h^-\pi^0$, for which the criterion $-58.3  < \Delta E < 35.0$~MeV  is applied. This $\Delta E$ criterion corresponds to a range of approximately three times the experimental resolution around zero. In addition, to suppress background from $D^{0}\to K^{0}_{\rm S}\pi^{0}$ in the $D^{0}\to\pi^{+}\pi^{-}\pi^{0}$ selection, requirements are placed on the vertex of the $\pi^{+}\pi^{-}$ pair to be consistent with
originating from the $e^{+}e^{-}$ collision point.  Figure~\ref{fig:hhpi0_signal_mbc} shows the $M_{bc}$ distributions for $CP$-tagged signal candidates, summed over all $CP$-even and $CP$-odd tags, respectively, where the $CP$-tag final state does not contain a $K^{0}_{\rm L}$ meson.

\begin{figure} 
\begin{center}
\begin{tabular}{cc}
\includegraphics[width=0.45\columnwidth]{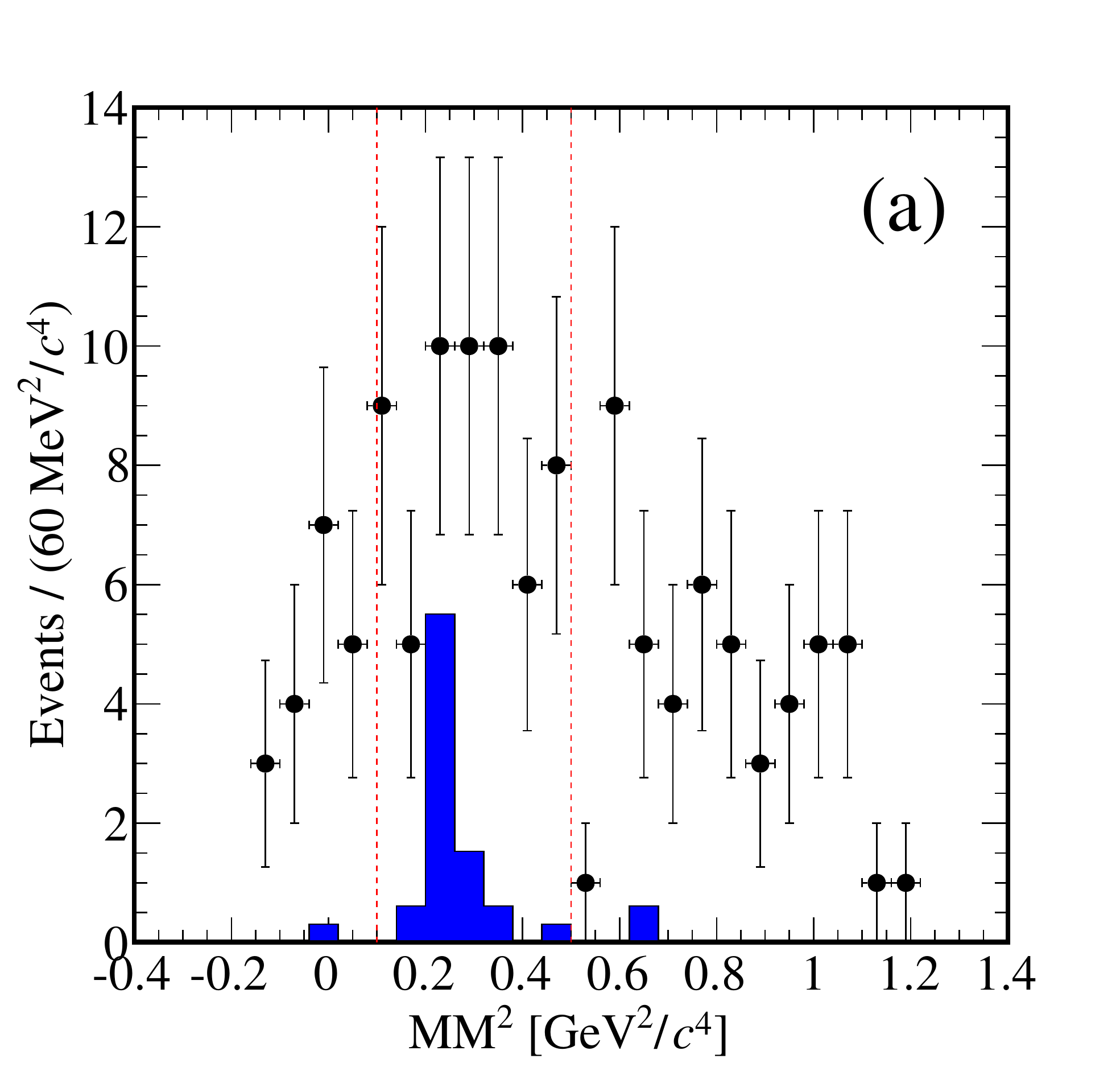} &
\includegraphics[width=0.45\columnwidth]{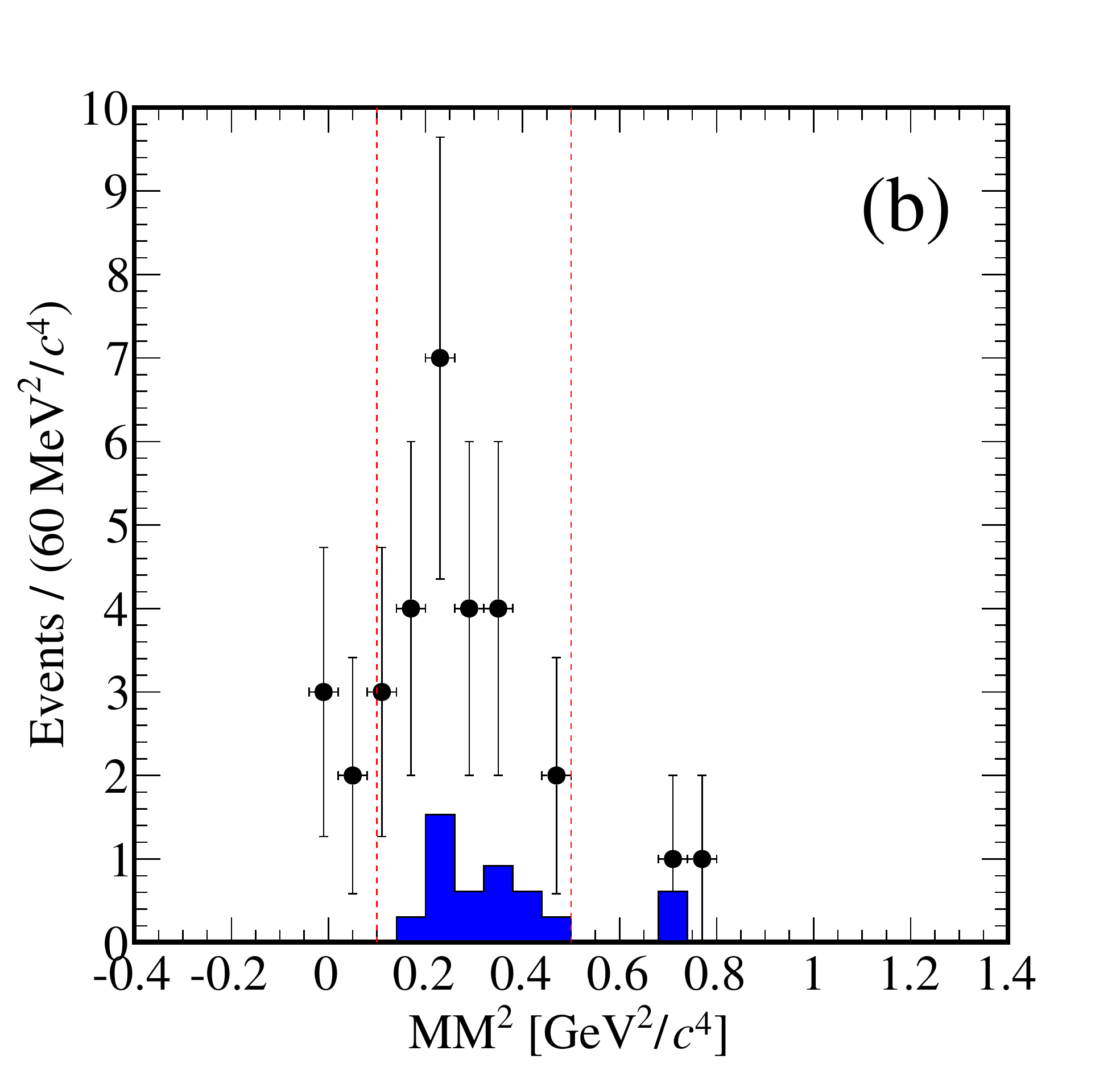}
\end{tabular}
\caption{$M_{\mathrm{miss}}^2$ distributions for $D \to\pi^{+}\pi^{-}\pi^{0}$ (a) and  $D\to K^{+}K^{-}\pi^{0}$ candidates tagged by $CP$ eigenstates that contain a $K^{0}_{\rm L}$. The shaded histogram indicates the peaking background. The vertical dotted lines indicate the applied signal window.} \label{fig:hhpi0_signal_MM2}
\end{center}
\end{figure}

 Many $K^{0}_{\rm L}$ mesons produced do not deposit any reconstructible signal in the detector. However, double-tag candidates can be fully reconstructed using a missing-mass squared $(M_{\mathrm{miss}}^2)$ technique \cite{K0LPRL} for tags containing a single $K^{0}_{\rm L}$ meson. Yields are extracted from the signal and sideband regions of the
$M_{\mathrm{miss}}^2$ distribution. Figure~\ref{fig:hhpi0_signal_MM2}  shows the $M_{\mathrm{miss}}^2$ distributions for candidates tagged with either a $K^0_{\rm L} \pi^0$ or $K^0_{\rm L} \omega$ tag.

\begin{figure} 
\begin{center}
\begin{tabular}{cc}
\includegraphics[width=0.45\columnwidth]{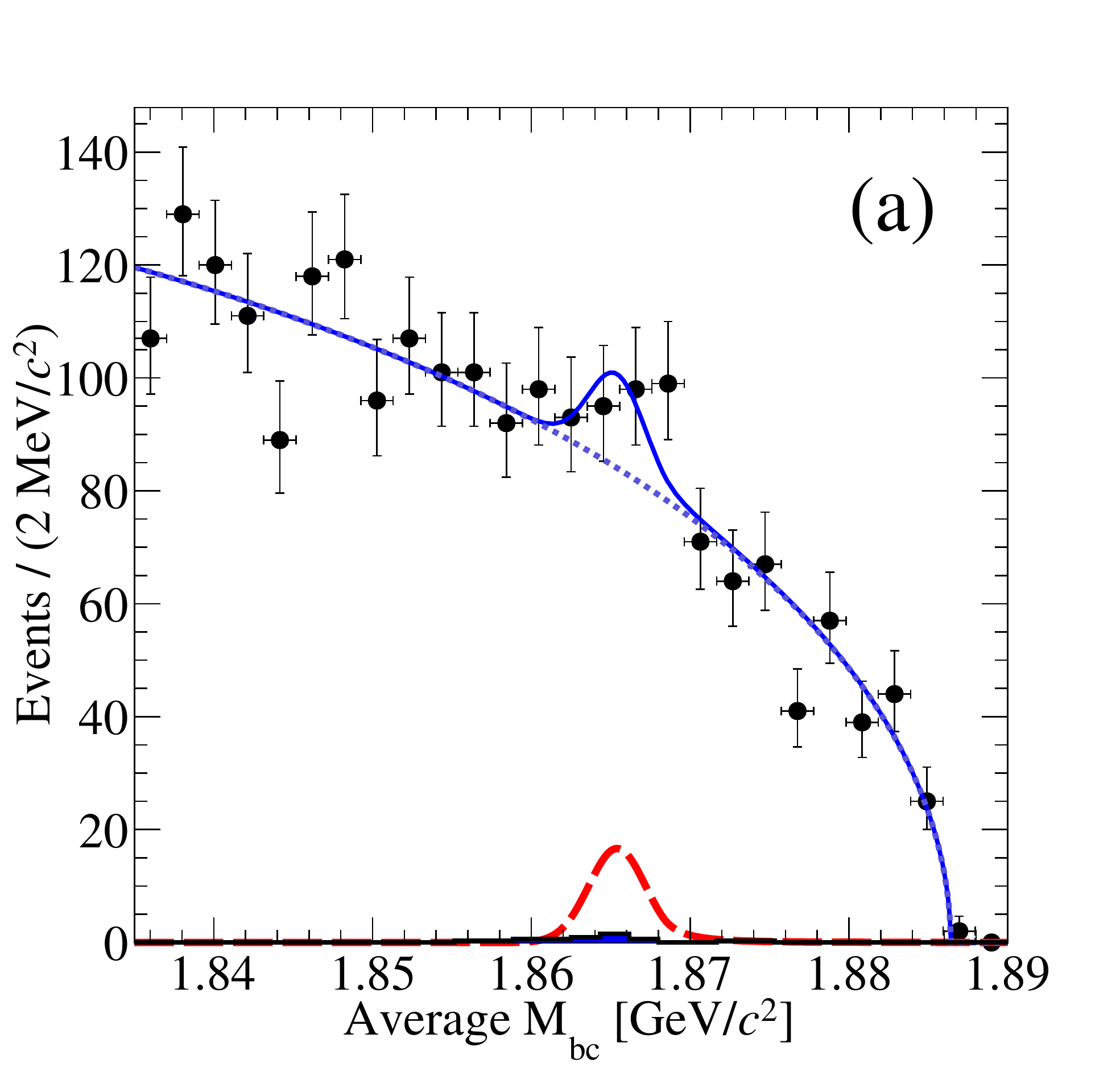} &
\includegraphics[width=0.45\columnwidth]{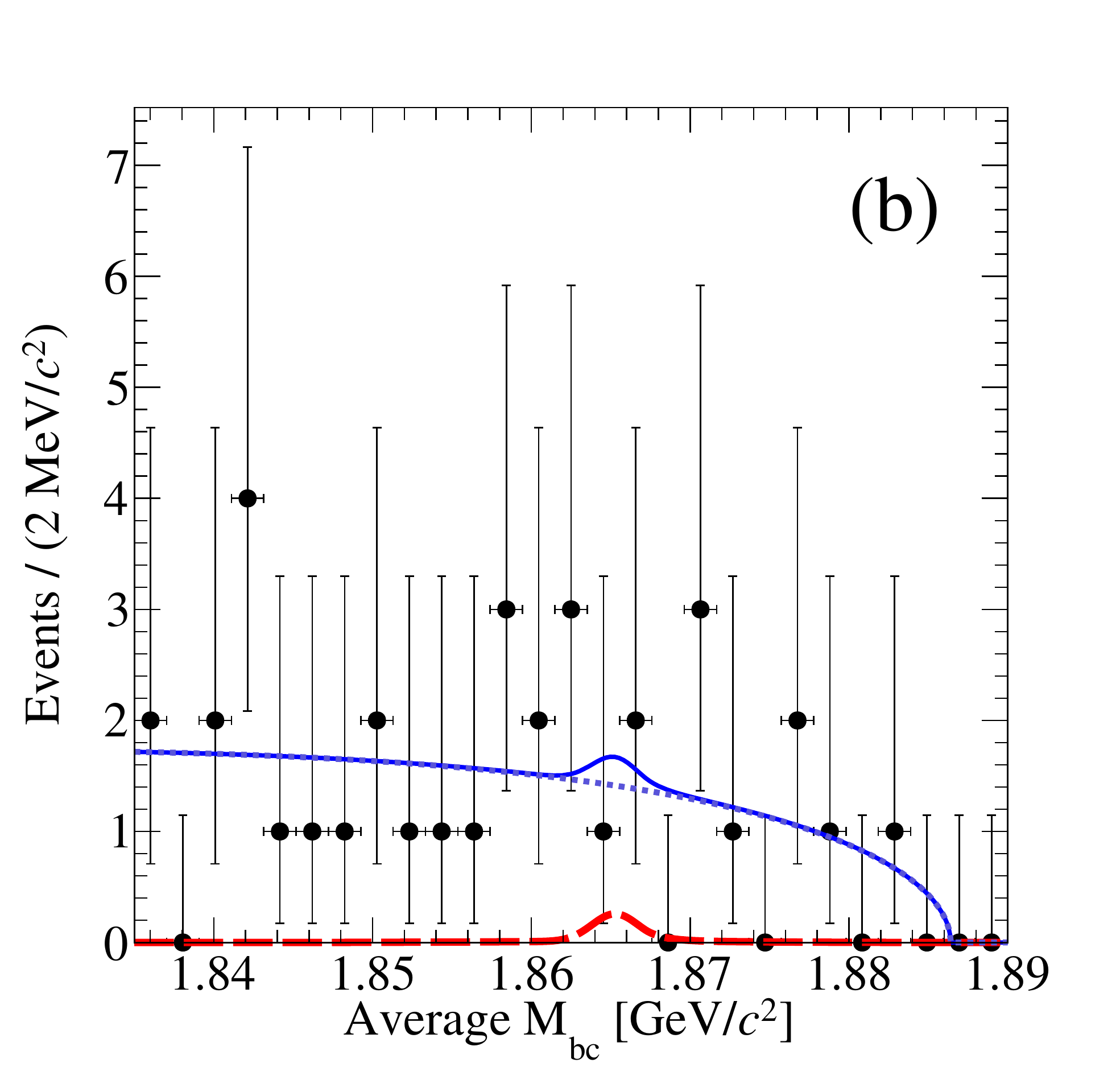}
\end{tabular}
\caption{Average $M_{bc}$ distributions (points with error bars) for (a) $\pi^+\pi^-\pi^0$ vs. $\pi^+\pi^-\pi^0$ and (b) $K^+K^-\pi^0$ vs. $K^+K^-\pi^0$. Superimposed are the total (solid line), signal (dashed line) and background (dotted line) fit results. The shaded histogram indicates the peaking background.} \label{fig:selftags}
\end{center}
\end{figure}

In the selection of $\pi^{+}\pi^{-}\pi^{0}$ vs. $\pi^{+}\pi^{-}\pi^{0}$ candidates an additional $K^0_{\rm S}$ veto is applied to remove $D \to K^0_{\rm S}(\pi^+\pi^-)\pi^0$ decays that are otherwise found to dominate the sample.  Candidates are rejected if they have a $\pi^+\pi^-$ invariant mass within 30~MeV/$c^2$ of the nominal $K^0_{\rm S}$ mass. The selected sample of $\pi^+\pi^-\pi^0$ vs. $\pi^+\pi^-\pi^0$ candidates contains a significant combinatoric background from continuum $e^{+}e^{-}\to u\bar{u},~ d\bar{d}$ events that hadronise to six pions. Furthermore, this combinatoric background does not follow a uniform distribution in $M_{bc}$ as in the other double-tag modes. Therefore, an alternative strategy is used to determine the signal yield. A maximum-likelihood fit to the distribution of the average $M_{bc}$ of the two $D\to\pi^+\pi^-\pi^0$ candidates is used to determine the signal yield. The probability density functions (PDFs) are parametrised by a Crystal Ball function \cite{CB} and a threshold ARGUS function \cite{ARGUS} for the signal and combinatoric background components, respectively. Apart from the signal yield all other  parameters of the signal PDF  are fixed to those obtained from the signal MC sample. All parameters for the background PDF are obtained from the fit to data. The average $M_{bc}$ distribution for $\pi^+\pi^-\pi^0$ vs. $\pi^+\pi^-\pi^0$ candidates is shown in Fig.~\ref{fig:selftags}(a), along with the result of the fit. No significant signal is observed. Even though the combinatoric background in the sample of $K^+K^-\pi^0$ vs. $K^+K^-\pi^0$ candidates is smaller, the same method is applied to determine the signal yield as for  $\pi^+\pi^-\pi^0$ vs. $\pi^+\pi^-\pi^0$. The average $M_{bc}$ distribution and fit result for $K^+K^-\pi^0$ vs. $K^+K^-\pi^0$ candidates is shown in Fig.~\ref{fig:selftags}(b); again no significant signal is observed. 

The peaking background estimates are determined from the generic MC sample of $D\overline{D}$ events. For final states without a $K^{0}_{\rm L}$ the peaking backgrounds are found to be extremely small or negligible. The peaking backgrounds are significant in the states tagged by $K^{0}_{\rm L}\pi^{0}$ and $K^{0}_{\rm L}\omega$ as shown in Fig.~\ref{fig:hhpi0_signal_MM2}. The dominant source of peaking background is $D^{0}\to K^{0}_{\rm S}X,~K^{0}_{\rm S}\to\pi^0\pi^{0}~(X=\pi^{0},\omega)$ events where the $\pi^{0}$ mesons from the $K^{0}_{\rm S}$ decay are not reconstructed. 

The measured event yields after background subtraction are given in 
Table~\ref{tab:yields}. No significant signal is seen in any of the modes tagged by a $CP$-even eigenstate, whereas significant signals are seen in most modes tagged by $CP$-odd eigenstates.

\begin{table}[th]
\begin{center}
\caption{Background subtracted signal yields and statistical uncertainties for double tags and the $CP$ single-tag yields.  } \vspace*{0.1cm}
\label{tab:yields}
\begin{tabular}{lccc}\hline\hline
Tag & $\pi^+\pi^-\pi^0$ & $K^+K^-\pi^0$  & Single \\ \hline
$\pi^+\pi^-\pi^0$ & $\phantom{0}34.3\pm20.0$  & -- & -- \\
$K^+K^-\pi^0$  & -- & $\phantom{0}2.7\pm 3.6$  & -- \\ \hline 
$K^{+}K^{-}$  & $\phantom{00}3.9\pm\phantom{0}5.5$  & $11.3\pm4.1$  & 11970 $\pm$ 116 \\
$\pi^{+}\pi^{-}$  & $\phantom{0}13.3\pm\phantom{0}6.6$ & $\phantom{0}1.7\pm3.8$ & \phantom{0}5595 $\pm$ 109 \\ 
$K^{0}_{\rm S}\pi^{0}\pi^{0}$  & $\phantom{00}1.9\pm\phantom{0}3.4$  & $\phantom{0}2.8\pm2.0$ &  \phantom{0}7306 $\pm$ 125 \\ 
$K^{0}_{\rm L}\pi^{0}$  & $\phantom{0}14.6\pm\phantom{0}7.9$ & $10.6\pm4.0$ & -- \\ 
$K^{0}_{\rm L}\omega$  & \hspace*{0.1cm}$-4.3\pm\phantom{0}4.0$ & $\phantom{0}3.5\pm4.4$ & -- \\ \hline
$K^{0}_{\rm S}\pi^{0}$   &$214.0\pm15.0$ & $26.0\pm5.2$ & 20069 $\pm$ 146 \\
$K^{0}_{\rm S}\omega$   &$\phantom{0}95.5\pm \phantom{0}9.9$ &  $11.6\pm3.4$ & \phantom{0}7960 $\pm$ \phantom{0}99 \\
$K^{0}_{\rm S}\eta(\gamma\gamma)$   & $\phantom{0}33.0\pm\phantom{0}5.8$  & $\phantom{0}3.5\pm2.4$ & \phantom{0}2903 $\pm$ \phantom{0}71 \\
$K^{0}_{\rm S}\eta(\pi^+\pi^-\pi^0)$   & $\phantom{00}8.8\pm\phantom{0}3.1$  & $\phantom{0}1.0\pm1.0$ & \phantom{0}1161 $\pm$ \phantom{0}48 \\
$K^{0}_{\rm S}\eta^{\prime}$  &$\phantom{0}17.6\pm\phantom{0}4.2$ & $\phantom{0}3.0\pm1.7$ & \phantom{0}1405 $\pm$ \phantom{0}38 \\ 
\hline\hline   
\end{tabular}
\end{center}
\end{table}

It is also necessary to know the single-tag yield for the $CP$-eigenstates to normalise the double-tagged yields appropriately to obtain a value of $F_{+}$. The selection of single tags is only possible for modes without a $K_{\rm L}^{0}$ in the final state. The selection criteria are identical to those for the double-tag selection. The signal yield is estimated using a maximum likelihood fit to the $M_{bc}$ distribution where the signal is modelled by the sum of a Gaussian and an asymmetric Gaussian and the combinatoric background is modelled by an ARGUS function. Apart from the signal yield all other  parameters of the signal PDF  are fixed to those obtained from the signal MC sample. All parameters for the background PDF are obtained from the fit to data. The signal yield is estimated by integrating the best-fit PDF within the interval $1.86 < M_{bc} < 1.87$~GeV/$c^2$.  Peaking backgrounds are estimated from the generic $D\bar{D}$ simulation. Significant contributions are only found for $D^{0}\to K^{0}_{\rm S}\omega$ and $D^{0}\to K^{0}_{\rm S} \eta^{\prime}(\pi^+\pi^-\pi^0)$ candidates, corresponding to 1.9\% and 3.8\% of the signal, respectively; in both cases the dominant sources of peaking background are states with intermediate $K^*$, $K_1$ and $K_2$ resonances which lead to  $K^{0}_{\rm S}\pi^+\pi^-\pi^0$ in the final state. The background-subtracted single-tag yields are given in Table~\ref{tab:yields}. 
\clearpage

\section{Results, systematic uncertainties and consistency checks}\label{sec:results}

The yields of double-tagged and single $CP$-tag candidates are used to determine the quantities $N^+$ and $N^-$, and from these the $CP$ fraction $F_+$.  The values for $N^+$ and $N^-$ are calculated from the ensemble of $CP$-odd and $CP$-even tags, respectively, accounting for statistical and systematic uncertainties, and allowing for the correlations that exist between certain systematic components.

There is an uncertainty in the single-tag yields $S^\pm_{\rm meas}$ associated with the fit function used to model the $M_{bc}$ distribution of the signal.  This shape of the distribution varies depending on whether there are no electromagnetic neutral final-state particles present ($K^+K^-$ and $\pi^+\pi^-$), whether the neutrals are relatively hard  ($K^0_{\rm S}\pi^0 (\gamma \gamma)$ and $K^0_{\rm S} \eta (\gamma \gamma)$) or soft (all other modes).  Uncertainties are assigned of $2.0\%$, $2.5\%$ and $5.0\%$, respectively. These assignments also adequately cover those uncertainties related to the assumption of the double-tag efficiency factorising into the product of the two single-tag efficiencies.
$S^\pm_{\rm meas}$ is corrected for the effects of $D^0\bar{D}^0$ mixing using $y_D = 0.62 \pm 0.08$~\cite{HFAG}.

Tags involving a $K^0_{\rm L}$ require special treatment as it is not possible to measure a single-tag yield for these modes.  The expected value   for the tag $K^0_{\rm L} \pi^0$ without mixing effects, $S^- (K^0_{\rm L}\pi^0)$,  is given by $2 \, N_{D\bar{D}}\, \epsilon_{K^0_{\rm L} \pi^0}\, \mathcal{B}_{h^+h^-\pi^0}$.  Here 
% $\epsilon_{K^0_{\rm L} \pi^0} = \epsilon_{K^0_{\rm L} \pi^0 | h^+h^-\pi^0} / \epsilon_{h^+h^-\pi^0}$ is an effective single tag
$\epsilon_{K^0_{\rm L} \pi^0}$ is an effective single tag 
efficiency, taken to be equal to the ratio of the double-tagged efficiency to the single-tagged signal efficiency, as determined from simulation and $ \mathcal{B}_{h^+h^-\pi^0}$ is the $D\to h^{+}h^{-}\pi^{0}$ branching fraction \cite{PDG}. The number of $D\bar{D}$ pairs in the sample, $ N_{D\bar{D}}$, can be measured from the double-tagged yield of decays into Cabibbo-favoured final states.  It is found that $S^-(K^0_{\rm L} \pi^0) = 24433 \pm 3934$, where the assigned error reflects the uncertainties in the input factors and assumptions of this calculation.  A similar procedure for $K^0_{\rm L}\omega$ yields $S^-(K^0_{\rm L} \omega) = 8923 \pm 4015$.

Finally, there is a possible source of bias arising from non-uniformities in the Dalitz acceptance.  The efficiency of reconstruction at CLEO-c is rather flat across phase space, but residual variations are parameterised and used to weight the amplitude models for the two signal modes, and the resulting effective values of $F_+$ are then calculated.  The potential bias is assessed to be $0.001$ for $\pi^+\pi^-\pi^0$, and $0.010$  for $K^+K^-\pi^0$. The significant  difference between the two values is attributed to the larger fraction of events in the $CP$-odd Dalitz plot for $D\to K^{+}K^{-}\pi^{0}$, which will be distributed differently than those in the  $CP$-even Dalitz plot. Therefore, the measured value of $F_{+}$ is affected more significantly by efficiency variations than for  $D\to \pi^{+}\pi^{-}\pi^{0}$.

\begin{figure}[th]
\begin{center}
\begin{tabular}{cc}
\includegraphics[width=0.45\columnwidth]{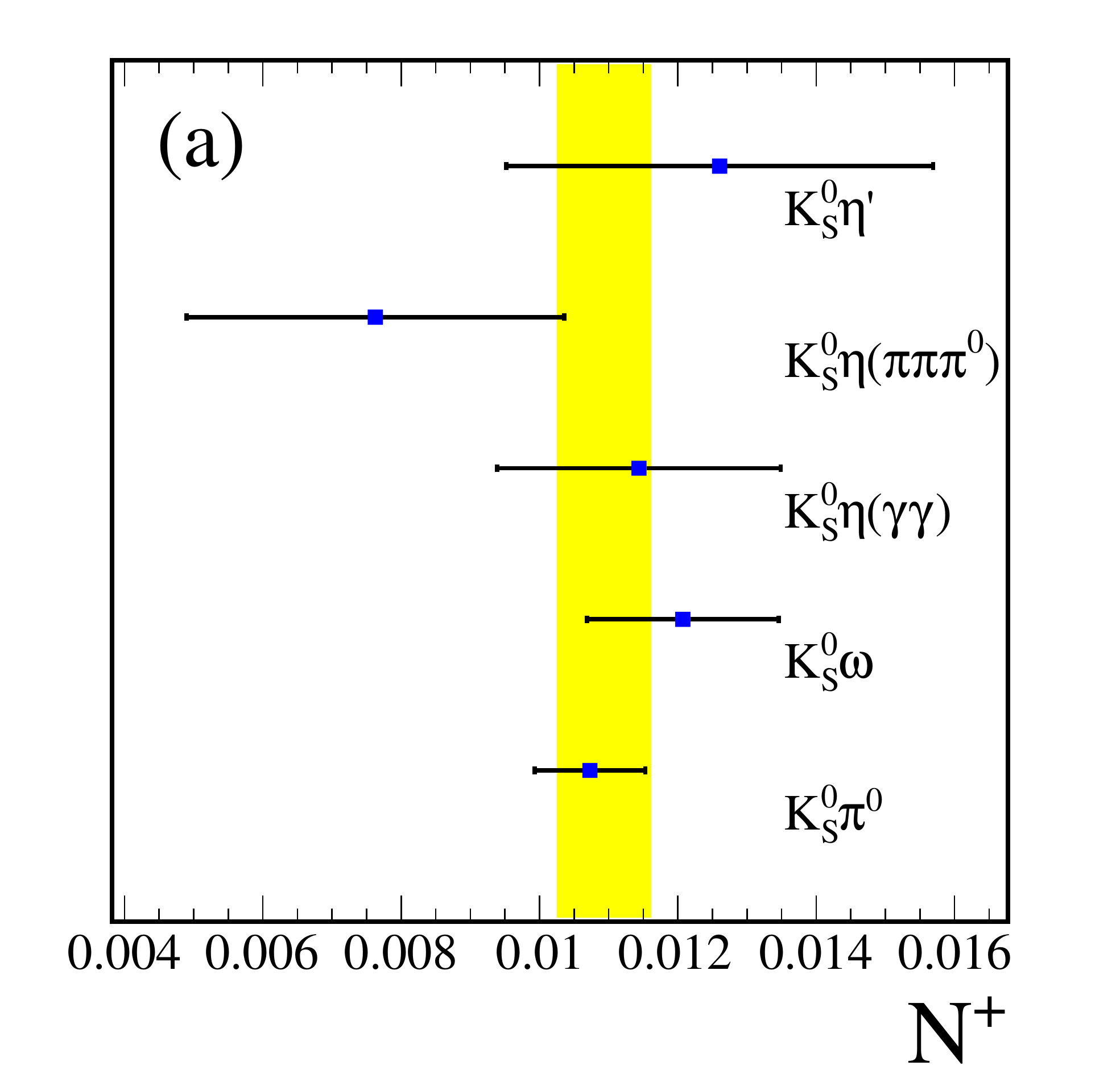} &
\includegraphics[width=0.45\columnwidth]{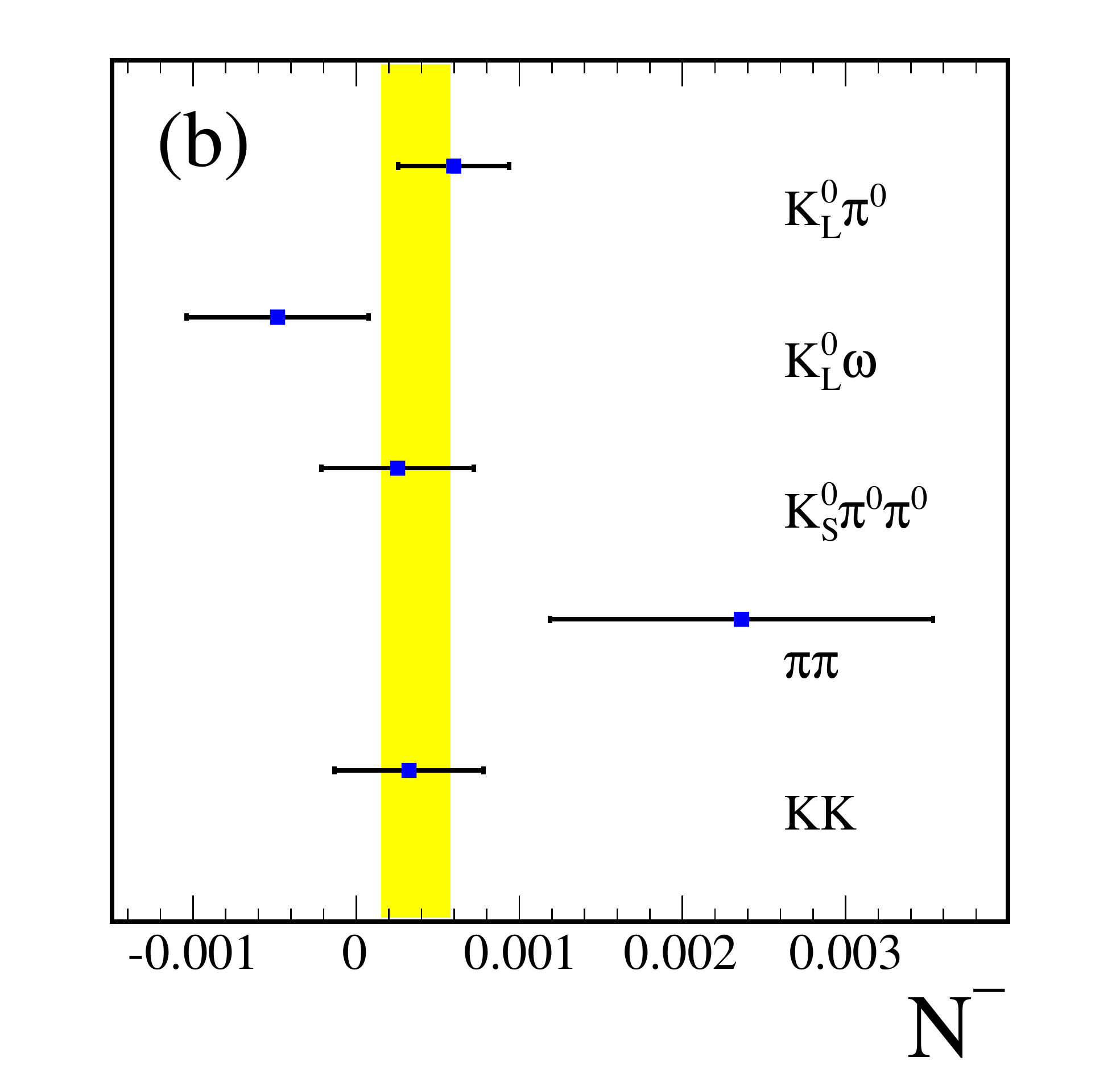} \\
\includegraphics[width=0.45\columnwidth]{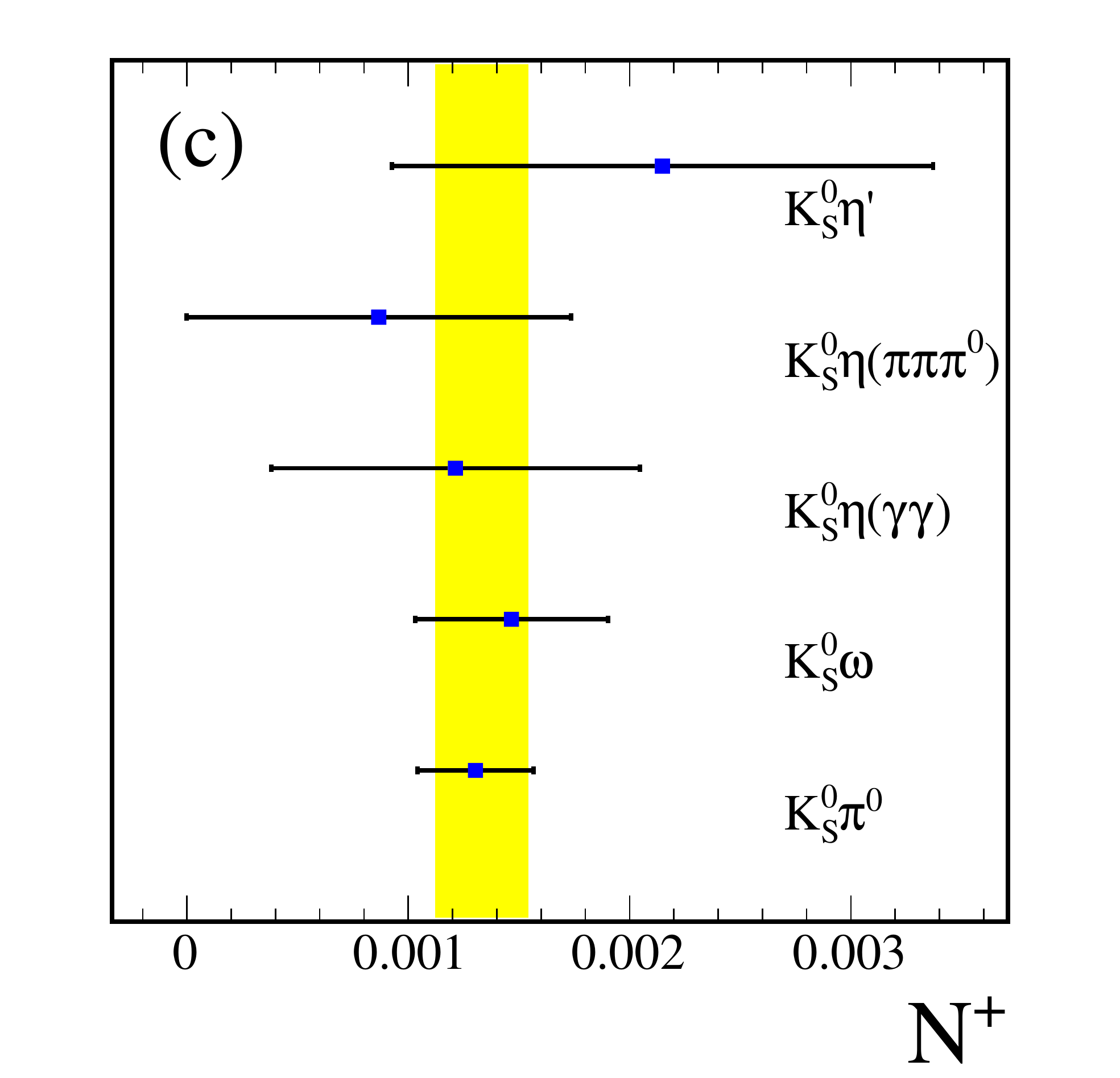} &
\includegraphics[width=0.45\columnwidth]{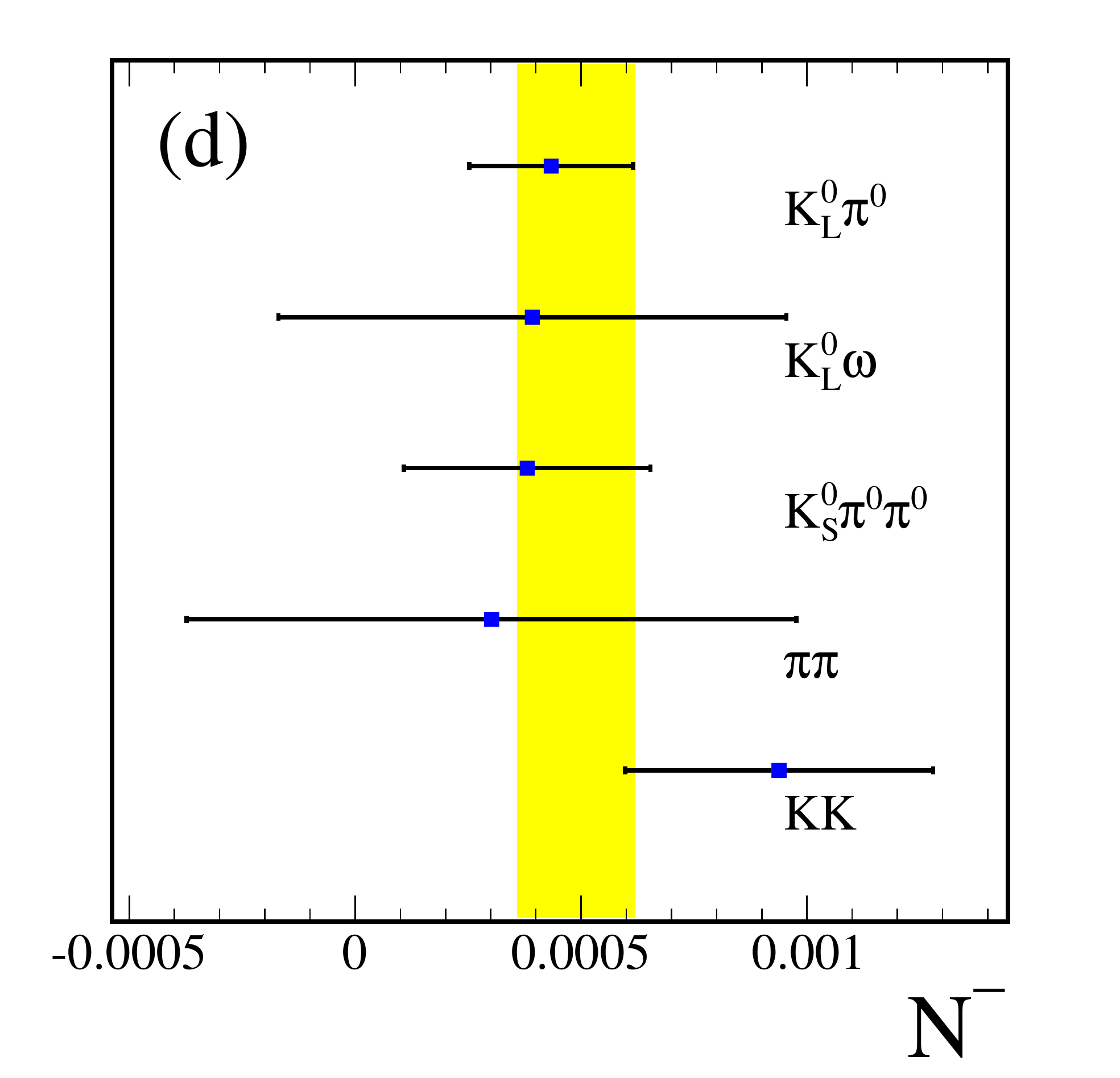} 
\end{tabular}
\caption{$D \to \pi^+\pi^-\pi^0$ results for $N^+$ (a) and $N^-$ (b).  $D \to K^+K^-\pi^0$ results for $N^+$ (c) and $N^-$ (d).
In each plot the vertical yellow band indicates the value obtained from the combination of all tags.  }  \label{fig:results}
\end{center}
\end{figure}

The measured values for $N^+$ and $N^-$ for the two signal modes are displayed in Fig.~\ref{fig:results}.  It can be seen that there is consistency between the individual tags for each measurement.  From these results it is determined that
$F_+ ({\pi^+\pi^-\pi^0}) = 0.968 \pm 0.017 \pm 0.006$  and  $F_+({K^+K^-\pi^0}) = 0.731 \pm 0.058 \pm 0.021$, where the first uncertainty is statistical and the second is systematic.   These values are slightly higher than, but compatible with, the model predictions  reported in Sect.~\ref{sec:cpcontent}.
%
% Here are results without efficiency systematic: 6/10/2014
%Pipip0:  0.968 +/- 0.017 +/- 0.0061
%KKpi0 :  0.731 +/- 0.058 +/- 0.0190
%

The results for the number of self tags are not used in the determination of the $CP$-even content because they provide much weaker constraints  than the $CP$-tags, and because the $K^0_{\rm S}$ veto imposed on the invariant mass of $\pi^+\pi^-$ pairs for the  $\pi^+\pi^-\pi^0$ vs.  $\pi^+\pi^-\pi^0$ selection distorts the Dalitz space for this mode.  Instead, the self-tagged yields are employed as a cross-check.
Making use of Eq.~\ref{eq:self} and taking the above values for the  $F_+ ({\pi^+\pi^-\pi^0})$ and  $F_+ ({\pi^+\pi^-\pi^0})$ leads to an expectation of $2.6 \pm 1.6$ and $0.4 \pm 0.1$ candidates  for the number of $D \to \pi^+\pi^-\pi^0$ 
and $D \to K^+K^-\pi^0$ self tags, respectively.  These predictions are compatible with the measured values reported in Table~\ref{tab:yields}.

\section{Implications for the measurement of  $\boldmath \gamma$} 
\label{sec:implications}

Sensitivity to the unitarity triangle angle $\gamma$ is obtained by measuring the relative rates of $B^\mp \to D({h^+h^-\pi^0}) K^\mp$ decays and related observables.  Once more considering the Dalitz plot to be divided into a pair of symmetric bins, as introduced in Sect.~\ref{sec:cpcontent}, and making use of the relations of Ref.~\cite{ANTONBONDAR} and  Eq.~\ref{eq:fplusbondar}, it follows that
\begin{eqnarray}
\Gamma(B^\mp \to D({h^+h^-\pi^0}) K^\mp ) & =  & h_B {\Big(} [ 1 +  r_B^2 ] [ 1  -  2 c_1 y_D \sqrt{K_1 K_{-1}} ]  \nonumber \\
&  & \: \: \: \: \: \: \: \: + \,  2x_\mp [ 2c_1 \sqrt{K_1 K_{-1}}  -  y_D ]  {\Big )} \nonumber \\
& = & h_B {\Big (}  [ 1 \, + \, r_B^2 ] [ 1 -  (2F_+ -1) y_D ]  \nonumber \\
&  & \: \: \: \: \: \: \: \: + \,  2x_\mp [ (2F_+ -1 )  - y_D  ]  {\Big ) \; .} 
% Old expressions without mixing
%\Gamma(B^\mp \to D({h^+h^-\pi^0}) K^\mp ) & =  & h_B ( 1 \, + \, r_B^2 \, + \, 4c_1 \sqrt{K_1 K_{-1}} x_\mp ) \nonumber \\
% &  = & h_B ( 1 \, + \, r_B^2 \, + \, 2 (2F_+ -1 ) x_\mp ).
\label{eq:bpartial}
\end{eqnarray}
Here $h_B$ is a normalisation factor, $r_B$ is the ratio of the magnitudes of the $B^+ \to D^0 K^+$ and $B^+ \to \bar{D}^0  K^+$ amplitudes, $\delta_B$ is the strong-phase difference between these amplitudes and $x_\pm = r_B \cos (\delta_B \pm \gamma)$. This expression includes the effects of $D^0\bar{D}^0$ oscillations at leading order in the mixing parameters~\cite{RAMA}.

These partial widths and those involving flavour-specific $D$ meson decays can be used to construct the  partial-widths ratio $R_{F_+}$ and $CP$-asymmetry $A_{F_+}$:
\begin{eqnarray}
R_{F_+}  & \equiv & 
\frac{ \Gamma(B^-\to D_{F_+} K^- ) \,+\,  \Gamma(B^+ \to D_{F_+} K^+ ) }
{ \Gamma(B^-\to D^0 K^- ) \,+ \,  \Gamma(B^+ \to \bar{D}^0 K^+ ) }, \\
A_{F_+} & \equiv & 
\frac{ \Gamma(B^-\to D_{F_+} K^- ) \,-\,  \Gamma(B^+ \to D_{F_+}K^+) }
{ \Gamma(B^-\to D_{F_+} K^- ) \,+ \,  \Gamma(B^+ \to D_{F_+ } K^+)} , 
\end{eqnarray}
where $D_{F_+}$ indicates a $D$ meson of $CP$-even content $F_+$, established through its decay into the final state $h^+h^-\pi^0$.  These observables are directly analogous to the usual so-called GLW~\cite{GLW} observables  $R_{{\rm CP }\pm}$ and $A_{{\rm CP} \pm}$, where the $D$ meson is reconstructed in a pure $CP$ eigenstate.\footnote{Experimentally, and following the usual procedure established in GLW analyses for measuring $R_{\rm CP \pm}$~\cite{BABARGLW,BELLEGLW,CDFGLW,LHCBGLW},  it is more convenient to determine $R_{F_+}$ from the double ratio $R_{F_+} \approxeq
\frac{ \Gamma(B^-\to D_{F_+} K^- ) \,+\,  \Gamma(B^+ \to D_{F_+}K^+)}
{ \Gamma(B^-\to D_{F_+} \pi^- ) \,+\,  \Gamma(B^+ \to D_{F_+}\pi ^+)} \, / \,
\frac{ \Gamma(B^-\to D_{K^-\pi^+} K^- ) \,+\,  \Gamma(B^+ \to D_{K^+\pi^-} K^+)}
{ \Gamma(B^-\to D_{K^-\pi^+} \pi^- ) \,+\,  \Gamma(B^+ \to D_{K^+\pi^-}\pi ^+)}$, where in the second ratio the $D$ mesons  are reconstructed in the Cabibbo-favoured decay.}

  In order to make explicit the relationship to the pure $CP$-eigenstate case, the effects of mixing are now neglected.  Then  $R_{F_+}$ and $A_{F_+}$ are found to have the following dependence on the underlying physics parameters:
\begin{eqnarray}
R_{F_+}  & = & 1 \, + \, r_B^2 + (2F_+ -1) \cdot 2r_B\cos\delta_B \cos \gamma,  \\
A_{F_+} &=& (2F_+ -1 ) \cdot 2r_B \sin\delta_B \sin\gamma / R_{{F_+}},
\end{eqnarray}
which reduces to the equivalent expressions for  $R_{{\rm CP }\pm}$ and $A_{{\rm CP} \pm}$ in the case $F_+$ is $1$ or $0$.  Therefore inclusive final states such as $h^+h^-\pi^0$ may be cleanly interpreted in terms of $\gamma$ and the other parameters of interest, provided that $F_+$ is known.  At leading order the only difference that the $CP$ asymmetry $A_{F_+}$ has with respect to the pure $CP$-eigenstate case is a dilution factor of $(2F_+ -1)$, which is $0.936 \pm 0.036$ for $D \to \pi^+\pi^-\pi^0$ and $0.462 \pm 0.124$ for  $D \to K^+K^-\pi^0$. The measurement of $F_{+}$ presented here assumes a uniform acceptance across the Dalitz plot; any non-uniformity is considered as a potential source of systematic uncertainty. Therefore, any non-uniformity of the acceptance over the Dalitz plot for an experiment determining $R_{F_+}$ and $A_{F_+}$ must be corrected for, if necessary, and a suitable systematic uncertainty assigned.

\section{Conclusion}\label{sec:conc}

Data corresponding to an integrated luminosity of 818~$\rm pb^{-1}$ collected by the CLEO-c experiment in $e^+e^-$ collisions at the $\psi(3770)$ resonance have been analysed for the decays $D \to \pi^+\pi^-\pi^0$ and $D \to K^+K^-\pi^0$.  Measurements of  $F_+$, the fractional $CP$-even content of each decay have been performed.  Values of $F_+ ({\pi^+\pi^-\pi^0}) = 0.968 \pm 0.017 \pm 0.006$  and  $F_+({K^+K^-\pi^0}) = 0.731 \pm 0.058 \pm 0.021$ are obtained, where the first uncertainty is statistical, and the second is systematic.
It has been demonstrated that such self-conjugate inclusive channels can be cleanly included in measurements of the unitarity-triangle angle $\gamma$, using $B^\mp \to D K^\mp$ decays.  The high value of $F_+$  obtained for  $D \to \pi^+\pi^-\pi^0$ makes this channel, in particular, a valuable addition to the suite of $D$-decay modes used in the measurement of $\gamma$ at LHCb and Belle-II. Furthermore, given the large branching fraction, the $D\to \pi^{+}\pi^{-}\pi^{0}$ state can provide an additional source of $CP$-even tags for quantum-correlated measurements at the $\psi(3770)$. The sample of $D\to\pi^{+}\pi^{-}\pi^{0}$ tags would be approximately twice as large as the $D\to h^{+}h^{-}$ sample. However, the formalism needs to be adjusted to incorporate $F_{+}$ to account for the small $CP$-odd component in the final state. Improved precision on the $F_+$ parameters can be obtained using the larger $\psi(3770)$ data set available at BESIII, and similar measurements can also be performed for other self-conjugate final states.

%\begin{figure}
%\centering
%\includegraphics[width=0.48\textwidth]{KsPiPi8binsForCharmMixing_smaller.pdf}
%\includegraphics[width=0.48\textwidth]{KsKK4binsForCharmMixing_smaller.pdf}
%\caption{Equal  $\Delta \delta_D$ BaBar binning for \KsPiPi (left) and for \KsKK (right).}
%\label{fig:binplots}
%\end{figure}

\section*{Acknowledgments}

This analysis was performed using CLEO-c data, and as members of the former CLEO collaboration we thank it for this privilege.
We  thank Brian Meadows and Kalanand Mishra for useful advice, and for helping us implement the BaBar amplitude models.
We have benefitted from valuable discussions with Mateo Rama, Anton Poluektov, Bhubanjyoti Bhattacharya and Jonathan Rosner. 
We are grateful for support from the UK Science and Technology Facilities Council and the UK-India  Education and Research Initiative.





\end{document}